%% file: main.tex
\documentclass[12pt,numbers,sort&compress]{article}
\usepackage{natbib}
\usepackage{hypernat}
\usepackage{amsmath}
\usepackage{amsthm}
\usepackage{amsfonts}          
\usepackage{amssymb}           
\usepackage[all]{xy}           
\usepackage{color}
\usepackage{graphicx}

\usepackage[colorlinks=true,linktocpage]{hyperref}

\usepackage[active]{srcltx}

\newlength{\xtrawidth}
\newlength{\xtraheight}
\newcommand{\eqdef}{%
  \mathrel{\lower.1mm
    \hbox{$\stackrel{\lower.424ex\hbox{\scriptsize def}}{=}$}}
}
\newcommand{\Z}{\mathbb{Z}}
\newcommand{\IR}{\mathbb{R}}
\newcommand{\C}{\mathbb{C}}
\newcommand{\CP}{\mathbb{P}}
\newcommand{\WP}{\mathbb{WP}}
\newcommand{\sufrak}{\mathfrak{su}}
\newcommand{\ufrak}{\mathfrak{u}}
\newcommand{\ptset}{\ensuremath{\{\text{pt.}\}}}
\newcommand{\modop}{~\mathrm{mod}~}

\DeclareMathOperator{\diag}{diag}

\DeclareMathOperator{\rank}{rank}
\DeclareMathOperator{\rk}{rk}

\DeclareMathOperator{\img}{img}
\DeclareMathOperator{\lcm}{lcm}

\DeclareMathOperator*{\bigtimes}{\hbox{\LARGE$\times$}}
\DeclareMathOperator{\degree}{deg}

\newcommand{\lrstack}[3][t]{
  \ensuremath{
    \begin{array}[#1]{c}
      \multicolumn{1}{l}{\displaystyle{#2}\quad} \\[0.5em] 
      \multicolumn{1}{r}{\displaystyle\quad{#3}}
    \end{array}
  }}

\def\clap#1{\hbox to 0pt{\hss#1\hss}}
\def\mathllap{\mathpalette\mathllapinternal}
\def\mathrlap{\mathpalette\mathrlapinternal}

\def\mathllapinternal#1#2{%
\llap{$\mathsurround=0pt#1{#2}$}}
\def\mathrlapinternal#1#2{%
\rlap{$\mathsurround=0pt#1{#2}$}}

\newcommand{\SA}{{\ensuremath{S^1_A}}}
\newcommand{\SB}{{\ensuremath{S^1_B}}}
\newcommand{\tK}[1][t]{{\ensuremath{{}^{#1}\!K}}}
\newcommand{\tL}[1][t]{{\ensuremath{{}^{#1}\!\mathcal{L}}}}
\newcommand{\tH}[1][t]{{\ensuremath{{}^{#1}\!H}}}

\newcommand{\GSO}{{\ensuremath{G_\text{GSO}}}}
\newcommand{\R}[1]{{\ensuremath{R{#1}}}}
\newcommand{\RZ}{{\R{\Z}}}
\newcommand{\RZt}{{\ensuremath{\widetilde{\RZ}}}}
\newcommand{\p}[2]{\ensuremath{p_{#1}^{#2}}}
\newcommand{\Kc}{\ensuremath{\mathcal{K}}}
\newcommand{\Hilb}{\mathcal{H}}

\begin{document}

\begin{titlepage}
  \begin{flushright}
    hep-th/0511100
    \\
    DESY-05-227
    \\
    UPR-1137-T
    \\
    ZMP-HH/05-20
  \end{flushright}
  \vspace*{\stretch{1}}
  \begin{center}
     \Huge 
      D-Brane Charges in Gepner Models
  \end{center}
  \vspace*{\stretch{2}}
  \begin{center}
    \begin{minipage}{\textwidth}
      \begin{center}
        \large         
        Volker Braun$^{\, \star}$ and 
        Sakura Sch\"afer-Nameki$^{\,\sharp}$
      \end{center}
    \end{minipage}
  \end{center}
  \vspace*{1mm}
  \begin{center}
    \begin{minipage}{\textwidth}
      \begin{center}
        ${}^\star$
        Department of Physics and Department of Mathematics\\        
        David Rittenhouse Laboratory, University of Pennsylvania\\
        209 S. 33rd Street, Philadelphia, PA 19104, USA
      \end{center}
      \vspace*{1mm}
      \begin{center}
        $^\sharp$
        II. Institut f\"ur Theoretische Physik 
        der Universit\"at Hamburg\\
        Luruper Chaussee 149, 22761 Hamburg, Germany\\
        and\\
        Zentrum f\"ur Mathematische Physik, 
        Universit\"at Hamburg\\
        Bundesstrasse 55, 20146 Hamburg, Germany
      \end{center}
    \end{minipage}
  \end{center}
  \vspace*{\stretch{1}}
  \begin{abstract}
    \normalsize 
    We construct Gepner models in terms of coset conformal field
    theories and compute their twisted equivariant K-theories. These
    classify the D-brane charges on the associated geometric
    backgrounds and therefore agree with the topological K-theories.
    We show this agreement for various cases, in particular the Fermat
    quintic.
  \end{abstract}
  \vspace*{\stretch{5}}
  \begin{minipage}{\textwidth}
    \underline{\hspace{5cm}}
    \centering
    \\
    Email: 
    \texttt{vbraun@physics.upenn.edu},
    \texttt{sakura.schafer-nameki@desy.de}.
  \end{minipage}
\end{titlepage}

\tableofcontents

\section{Introduction}
\label{sec:intro}

It is by now firmly established~\cite{MooreMinasian,
  WittenDbranesKtheory} that the K-theory groups of space-time are the
D-brane charge groups. More precisely, the claim is that the
isomorphism classes of D-brane boundary states modulo boundary
renormalization group (RG) flow are in one to one
correspondence~\cite{MooreKphysics} with suitable K-theory classes of
the string theory background in question. For geometrical backgrounds such as 
Calabi-Yau manifolds 
one can construct a variety of D-branes by applying methods from
boundary CFT, matrix factorizations and geometry 
\cite{Recknagel:1997sb, Recknagel:2002qq, Recknagel:2003nk, 
Brunner:2005fv, Brunner:2005pq, Enger:2005jk, Fredenhagen:2005an, Caviezel:2005th}. However,  determining the
endpoint of the RG flow~\cite{FredenhagenSchomerus}
is unfortunately not easy.

Most well-understood in this context are purely geometrical
backgrounds of string theory, such as tori, orbifolds, and Calabi-Yau
manifolds. In these instances, the K-theories were either already
available in the mathematics literature or are easily computed by
standard techniques and the complementary string theory computation of
D-brane charges is relatively straightforward.

Less trivial is the situation of string theory backgrounds with
non-trivial NSNS three-form flux $H$, where it is believed that
twisted K-theory is the correct structure to classify D-brane
charges~\cite{WittenDbranesKtheory, BouwknegtMathai, Aussies}.
Explicit checks of this claim have so far been restricted to
backgrounds with large symmetries, namely supersymmetric WZW and coset
conformal field theories (CFT)s~\cite{MMSI, MMSII, VolkerLieK, Fredenhagen:2003xf, 
Schafer-Nameki:2003aj, Gaberdiel:2003kv,  Gaberdiel:2004yn, Braun:2004qg, FredenhagenSO3, Gaberdiel:2004hs, Gaberdiel:2004za,Schafer-Nameki:2004yr}. 
The computation of twisted K-theories for compact Lie groups and coset models thereof 
were greatly simplified by the theorem of Freed, Hopkins, and
Teleman~\cite{FreedVerlindeAlg, FHTcomplex, FHTintegral}.

The objective of this paper is to test the twisted K-theory proposal
beyond standard CFT backgrounds by extending it to Gepner models.
These are essentially orbifolds of tensor products of ${\cal N}=2$
minimal models, realized for our purposes in terms of $SU(2)/U(1)$
supersymmetric coset models. They are known to describe certain tori
and Calabi-Yau spaces at particular points in their moduli space.
Because the K-groups are a topological quantity, the D-brane charge
group should be independent of the moduli. Therefore the twisted
equivariant K-theory of the Gepner models has to agree with the
topological K-theory of the corresponding Calabi-Yau manifold. This
provides a non-trivial check of the brane charge classification.


Technically, we are going to make use of the twisted equivariant Chern
character. Consequently, we are going to compute the complexified
K-groups 
\begin{equation}
  K^\ast(X;\C) \eqdef K^\ast(X) \otimes_\Z \C
\end{equation}
only. The downside is that one looses interesting
torsion~\cite{Braun:2000zm, Brunner:2001eg, Brunner:2001sk}
information, since
\begin{equation}
  K^\ast(X) = \Z^r \oplus \Z_{n_1} \oplus \cdots \oplus \Z_{n_k}
  \quad \Rightarrow \quad
  K^\ast(X;\C) = \C^r
  \,.
\end{equation}
However, the since none of the Calabi-Yau threefolds with Gepner
points actually have torsion in their K-group we do not expect to find
any in the Gepner models either.

During the final stage of this work we received a
preprint~\cite{Caviezel:2005th} that constructs a basis of D-branes
for the D-brane charge group. We will discuss a few details of their
approach in Section~\ref{sec:mrg}.

\section{The Quintic}
\label{sec:quintic}

As an hors d'{\oe}uvre to our work, let us
discuss~\cite{Recknagel:1997sb, Brunner:1999jq} the $(k=3)^5$ Gepner
model.  It is know to correspond to the Fermat quintic
\begin{equation}
  Q = 
  \Big\{ 
  [x_0:x_1:x_2:x_3:x_4]
  \Big|
  \sum_{i=0}^4 x_i^5 = 0
  \Big\}
  \subset 
  \CP^4
  \,.
\end{equation}
The Hodge diamond of the quintic is by now quite familiar to all
string theorists, and reads
\begin{equation}
  h^{pq}(Q)=~
  \vcenter{\xymatrix@!0@=7mm@ur{
    1 &  0 &  0 & 1 \\
    0 & 101 & 1 & 0 \\
    0 & 1 & 101 & 0 \\
    1 &  0 &  0 & 1 
  }}
  \,.
\end{equation}
We also know what there is no torsion in its cohomology, which then
determines its K-theory to be 
\begin{equation}
  \begin{aligned}
    K^0(Q) =&~ 
    H^\text{even}(Q;\C) = 
    \Z^4
    && \Rightarrow &
    K^0(Q;\C) =&~ \C^4
    \,, \\
    K^1(Q) =&~  
    H^\text{odd}(Q;\Z) = 
    \Z^{204}
    && \Rightarrow &
    K^1(Q;\C) =&~ \C^{204}
    \,.
  \end{aligned}
\end{equation}
We are going to arrive at the same answer for the complexified
K-groups directly from the Gepner model, without making any reference
to the quintic hypersurface. 

The Gepner model corresponding to the quintic is a $\Z_5$ orbifold of
$5$ copies of the level $k=3$ minimal model, see
Sections~\ref{sec:coset} and~\ref{sec:kgroups} for more details.
Moreover, the minimal model can be realized as an
$\sufrak(2)_k/\ufrak(1)$ coset CFT. The coset CFT has a nice sigma
model interpretation, it is an $SU(2)$ WZW model with a gauged $U(1)$
action. More precisely, the $U(1)$ acts\footnote{The cognoscente of
  course realize that our choice of maximal torus $U(1)\subset SU(2)$
  is random. Since all maximal tori are conjugate, we just picked this
  one for explicitness.} as
\begin{multline}
  U(1) \times SU(2) \to SU(2)
  \,,\quad 
  \\
  \Big[ e^{i \theta}, 
  \begin{pmatrix}
    a & b \\ c & d 
  \end{pmatrix}
  \Big]
  ~\mapsto~
  \begin{pmatrix}
     \cos \frac{\theta}{2} & \sin \frac{\theta}{2} \\ 
    -\sin \frac{\theta}{2} & \cos \frac{\theta}{2}
  \end{pmatrix}  
  \begin{pmatrix}
    a & b \\ c & d 
  \end{pmatrix}
  \begin{pmatrix}
     \cos \frac{\theta}{2} & \sin \frac{\theta}{2} \\ 
    -\sin \frac{\theta}{2} & \cos \frac{\theta}{2}
  \end{pmatrix}^{-1}
\end{multline}
See also Figure~\ref{fig:orbitsA} for a picture of the orbits. The
fixed point set of the $U(1)$ action is a circle inside $SU(2)\simeq
S^3$, which we denote by\footnote{For any space $X$ with action of a
  group $G$, we write $X^G$ for the $G$-fixed points. If $g\in G$,
  then we write $X^g$ for the points fixed by the subgroup
  $\left<g\right> \subset G$.}
\begin{equation}
  \SA
  \eqdef 
  \big[ SU(2) \big]^{U(1)} 
  = 
  \left\{ 
  \begin{pmatrix}
     \cos \varphi & \sin \varphi \\
    -\sin \varphi & \cos \varphi
  \end{pmatrix}  
  \Big|
  \varphi \in [0,\dots, 2\pi)
  \right\}
  \,.
\end{equation}
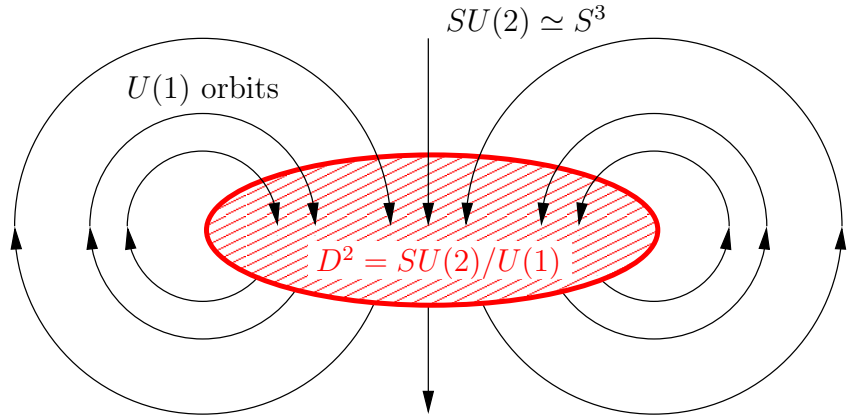
\begin{figure}[htbp]
  \centering
  \input{S1A.pstex_t}
  \caption{$U(1)$ action on $SU(2)$.}
  \label{fig:orbitsA}
\end{figure}
The space of orbits $SU(2)/U(1)$ is a disk, bounded by the fixed
points \SA. Rotating this disk is another symmetry of the geometry,
but arbitrary rotations are not a symmetry of the theory. The reason
is that the $H$ field is not symmetric under arbitrary rotations of
the disk. Rather, the rotation group is broken to rotations by
$\frac{2\pi}{5}$.
\begin{figure}[htbp]
  \centering
  \input{S1B.pstex_t}
  \caption{$\Z_5$ action on $SU(2)$.}
  \label{fig:orbitsB}
\end{figure}
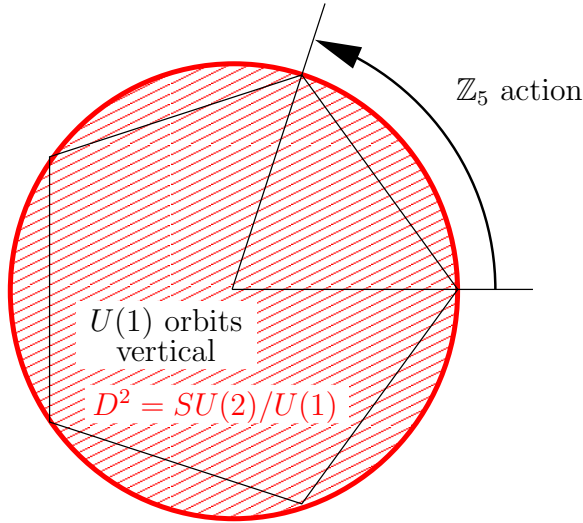
This $\Z_5$ group action lifts to an action on the
$SU(2)$ with fixed point set \SB, see figure~\ref{fig:orbitsB}. The
fixed point sets \SA{} and \SB{} form a Hopf link inside $SU(2)\simeq
S^3$. 

By now it is firmly established that the charge group is given by the
K-theory of space-time. More precisely, one has to pick the right
``flavor'' of K-theory depending on which $\mathcal{N}=1$
supersymmetric theory one formulates on the
background~\cite{Gaberdiel:2002jr, Schafer-Nameki:2003aj,
  Braun:2004qg, Schafer-Nameki:2004yr}. For the coset model, the
background is $SU(2)$ with an $H$ flux. The latter implies that the
correct K-theory is the so-called twisted K-theory, which we denote by
$\tK$. Moreover, we want to gauge a $U(1)$ symmetry. As is familiar to
all string theorists, this does \emph{not} mean that we work on the
set theoretic quotient $SU(2)/U(1)$. Instead, we have to correctly
incorporate the twisted sectors, which on the level of cohomology
means that we have to compute the $U(1)$ equivariant cohomology
groups. Therefore, the correct K-theory for the minimal model is
\begin{equation}
  \text{D-brane charges in $\sufrak(2)_k/\ufrak(1)$ coset}
  ~=~
  \tK_{U(1)}\Big( SU(2) \Big)
\end{equation}
with the twist class 
\begin{equation}
  t = k+2 ~\in~ \Z = H^3_{U(1)}\Big(SU(2);\Z\Big)
  \,.
\end{equation}
Hence the D-brane charges in the tensor product of $5$ minimal models
are
\begin{equation}
  \tK_{U(1)\times U(1)\times U(1)\times U(1) \times U(1)}
  \Big(
  SU(2)\times
  SU(2)\times
  SU(2)\times
  SU(2)\times
  SU(2)
  \Big)
  \,,
\end{equation}
where each $U(1)$ acts on just one of the $SU(2)$ factors. Finally,
the Gepner model is the $\Z_5$ orbifold by the diagonal $\Z_5$
action. Therefore
\begin{equation}
  \text{D-brane charges in the $(k=3)^5$ Gepner model}
  ~=~
  \tK_{U(1)^5\times \Z_5}\Big( SU(2)^5 \Big)
\end{equation}
To compute these K-groups we are using a twisted version of the
equivariant Chern isomorphism\footnote{In this paper, we are only
  concerned with Abelian groups $G$. In general the sum is over
  conjugacy classes.}
\begin{equation}
  ch:~
  K_G^{0,1}\big(X;\C\big) 
  \stackrel{\sim}{\longrightarrow}
  \bigoplus_{g\in G}
  H_G^{\text{even, odd}}\big(X^g;\C\big)
  \,.
\end{equation}
Adding an additional twist to the equivariant Chern character has two
consequences. First, one is lead to twisted cohomology, which is
roughly the cohomology of $d+[H]$ instead of $d$ on differential
forms. Second, the cohomology is with local coefficients, that is with
coefficients in a flat line bundle $\tL$ instead of the trivial flat
line bundle $\C$. The ensuing twisted equivariant Chern character (see
Section~\ref{sec:chern})
\begin{equation}
  \label{eq:twistequivchern}
  ch:~
  \tK_G^\ast\big(X;\C\big) 
  \longrightarrow
  \bigoplus_{g\in G}
  \tH_G^\ast\Big(X^g;\tL(g)\Big)  
\end{equation}
is an isomorphism, provided that only finitely many summands on the
right are non-vanishing. This turns out to be the case here, and
\begin{equation}
  \label{eq:twistedequivariantchern}
  \begin{split}
    \tK_{U(1)^5\times \Z_5}^\ast\big(SU(2)^5;\C\big) 
    \simeq&~
    \bigoplus_{g\in {U(1)^5\times \Z_5}}
    \tH_{U(1)^5\times \Z_5}^\ast\Big(\big[SU(2)^5\big]^g;\tL(g)\Big)  
    \\
    =&~
    \bigoplus_{g\in {U(1)^5\times \Z_5}}
    \bigg[
    \tH_{U(1)^5}^\ast\Big(\big[SU(2)^5\big]^g;\tL(g)\Big)  
    \bigg]^{\Z_5}    
  \end{split}
\end{equation}
is indeed an isomorphism. More specifically, as we are going to show
in Section~\ref{sec:chern} the only contributions are from the $4^5+4$
group elements
\begin{subequations}
  \begin{align}
    g =&~
    (\omega^{m_1},\omega^{m_2},\omega^{m_3},\omega^{m_4},\omega^{m_5},1)
    \,,\quad 
    m_i \in \{1,\dots,4\}
    \,,
    \\
    g =&~ (1,1,1,1,1,n) 
    \,,\quad 
    n\in\{1,\dots,4\}
  \end{align}  
\end{subequations}
in $U(1)^5\times\Z_5$, where we write $\omega\eqdef\exp(\frac{2\pi
  i}{5})$. The corresponding fixed point sets are of the form
\begin{subequations}
  \begin{align}
    g =&~
    (\omega^{m_1},\omega^{m_2},\omega^{m_3},\omega^{m_4},\omega^{m_5},1)
    & 
    \Rightarrow \Big[SU(2)^5\Big]^g =&~ (\SA)^5
    \,,
    \\
    g =&~ (1,1,1,1,1,n)
    & 
    \Rightarrow \Big[SU(2)^5\Big]^g =&~ (\SB)^5
    \,.
  \end{align}
\end{subequations}
As we are going to discuss in more detail in the next section, the
twisted equivariant cohomology for a single factor
$\tH_{U(1)}\big(SU(2)^g;\tL(g)\big)$ for $g\in U(1)\times\Z_5$ is 
\begin{subequations}
  \begin{align}
    g =&~
    (\omega^m,1)
    ~\Rightarrow\quad
    \tH_{U(1)}^0\big(\SA;\tL(g)\big) =
    0 
    \,, \quad 
    \tH_{U(1)}^1\big(\SA;\tL(g)\big) =
    \omega^m
    \,,
    \\
    g =&~ (1,n)
    ~\Rightarrow\quad
    \tH_{U(1)}^0\big(\SB;\tL(g)\big) =
    1
    \,, \quad 
    \tH_{U(1)}^1\big(\SB;\tL(g)\big) =
    0
    \,.
  \end{align}
\end{subequations}
where we write the cohomology groups as $\Z_5$ characters\footnote{By
  abuse of notation, we denote the generator for the character ring
  again $\omega$. In other words, $m\in\Z_5=\{0,\dots,4\}$ acts by
  multiplication with $\omega^m=\exp(\frac{2\pi i m}{5})$.}. The
cohomology groups for the tensor product of $5$ such factors is
readily determined from the K\"unneth formula, and one obtains\
\begin{subequations}
  \begin{align}    
    \bigoplus_{g = (\omega^{m_1},\dots,\omega^{m_5},1)}
    \tH_{U(1)^5}^\ast\Big((\SA)^5;\tL(g)\Big)  
    =& 
    \begin{cases}
      0
      \,, & 
      \ast=0 
      \,;
      \\
      \Big( \omega + \omega^2 + \omega^3 + \omega^4 \Big)^5
      \,, & 
      \ast = 5 \equiv 1\modop 2 
      \,;
    \end{cases}
    \\
    \bigoplus_{g = (1,\dots,1,\omega^n)}
    \tH_{U(1)^5}^\ast\Big((\SB)^5;\tL(g)\Big)  
    =& 
    \begin{cases}
      4
      \,, & 
      \ast=0 
      \,;
      \\
      0
      \,, & 
      \ast = 1
      \,.
    \end{cases}    
  \end{align}
\end{subequations}
It is now easy to determine the $\Z_5$-invariant part. Using the
twisted equivariant Chern character
eq.~\eqref{eq:twistedequivariantchern}, we obtain
\begin{equation}
  \label{eq:Kquintic}
  \tK_{U(1)^5\times \Z_5}^\ast\Big(SU(2)^5;\C\Big) 
  =
  \begin{cases}
    \C^{4}
    \,, &
    \ast = 0
    \,; \\
    \bigg[
    \Big( \omega + \omega^2 + \omega^3 + \omega^4 \Big)^5
    \bigg]^{\Z_5}
    =
    \C^{204}
    \,, &
    \ast = 1
  \end{cases}
\end{equation}
which precisely equals the K-theory of the quintic\footnote{Perhaps
  not surprisingly, formally the same computation arises when one
  tries~\cite{Kapustin:2004df} to construct Gepner models using matrix
  factorizations. However, the authors of~\cite{Kapustin:2004df} fail
  to address the twisted sector branes that arise when the Gepner
  model contains minimal models of different levels.} hypersurface.

\section{K-Theory of Gepner Models}
\label{sec:Theory}

\subsection{Group Theory}
\label{sec:group}

As we saw in the quintic example discussed in
Section~\ref{sec:quintic}, one has to determine cohomology groups
which form representations under a discrete group $\GSO$ (=$\Z_5$ for
the quintic) which implements the GSO\footnote{After Gliozzi, Scherk,
  and Olive~\cite{Gliozzi:1976qd}.} projection. Now we could always
work with polynomials of characters as in eq.~\eqref{eq:Kquintic}, but
this becomes cumbersome if one has to deal with tensor products of
different minimal models.

For cyclic groups $\Z_\kappa$, the following
representations\footnote{In this paper, we are only going to consider
  complex representations.} will appear again and again.
\begin{itemize}
\item The trivial representation $\C$.
\item The regular representation $\RZ_\kappa$, which is defined as
  follows: Take the vector space $\C^\kappa$. The group acts by
  cyclically permuting the $\kappa$ basis vectors. The regular
  representation can be diagonalized to the sum of all $1$-dimensional
  representations. Explicitly, if $\chi:\Z_\kappa\to\C$,
  $\chi(1)=\exp(\frac{2\pi i}{\kappa})$ is the generating character
  then
  \begin{equation}
    \RZ_\kappa = \bigoplus_{i=0}^{\kappa-1} \chi^i
    \,.
  \end{equation}
\item The representation $\RZt_\kappa$, which is the regular
  representation without its trivial sub-representation
  \begin{equation}
    \RZt_\kappa = \bigoplus_{i=1}^{\kappa-1} \chi^i
    \,.
  \end{equation}
  More formally, it is the cokernel
  \begin{equation}
    0 \longrightarrow
    \C \longrightarrow 
    \RZ_\kappa \longrightarrow
    \RZt_\kappa \longrightarrow
    0
  \end{equation}
\end{itemize}
Moreover, since we are actually computing cohomology groups everything
has a cohomological $\Z_2$-grade. By definition, we assign
\begin{subequations}
  \begin{align}
    \label{eq:degC}
    \degree\big( \C \big) =&~ 0 
    \,,
    \\
    \label{eq:degRZ}
    \degree\big( \RZ_\kappa \big) 
    =
    \degree\big( \RZt_\kappa \big) 
    =&~ 1
    \,.
  \end{align}
\end{subequations}

Of course, we have the usual operations of restriction and induction
(transfer) to relate $\GSO$-representations and representations of
subgroups of $\GSO$. However, $\GSO$ is always a cyclic group and we
have yet another operation which will occur frequently. This works as
follows. Given any subgroup $\Z_\kappa \subset \GSO$, we have in
addition to the inclusion $i$ also a projection $\pi$
\begin{equation}
  \vcenter{\xymatrix@C=3cm{
      \mathllap{
        \Z_\kappa 
      }
      \ar@<1mm>@/^3mm/[r]^-{i:~ n\mapsto \frac{|\GSO|}{\kappa}n}
      \ar@<-1mm>@{<-}@/_3mm/[r]_-{\pi:~ n\mapsto n\modop\kappa}
      & 
      \mathrlap{ 
        \GSO \simeq \Z_{|\GSO|} 
      }
  }}
\end{equation}
by modding out by $\kappa$. Given a representation $\rho:\Z_\kappa\to
\C^n$, we can then define a representation $\p{\Z_\kappa}{\GSO}(\rho)$
of $\GSO$ on the same vector space $\C^n$ by composing 
\begin{equation}
  \label{eq:pdef}
  \p{\Z_\kappa}{\GSO}(\rho)
  \eqdef 
  \rho \circ \pi
  :~
  \GSO \to \C^n
  \,,~
  (n,v) \mapsto \rho\big( n\modop\kappa, v\big)
  \,.
\end{equation}
Now in general the projection $\pi$ depends on which generators you
chose for $\GSO$, a random choice. However, for the identity, the
regular, and the reduced regular representation of $\Z_\kappa$ the
resulting $\GSO$ representation does \emph{not} depend on that choice.
We are only going to use the $\p{\Z_\kappa}{\GSO}$ operation in these
cases. 

For example, consider the group $\Z_{12}=\{0,1,\dots,11\}$ with the
character $\chi(1)=e^{\frac{2\pi i }{12}}$. Then the representation
\begin{multline}
  \p{\Z_3}{\Z_{12}}\Big( \RZt_3 \Big) 
  ~\otimes_\C~
  \p{\Z_4}{\Z_{12}}\Big( \RZt_4 \Big)
  = \\ =
  \Big( \chi^4 + \chi^8 \Big)
  \Big( \chi^3 + \chi^6 + \chi^9 \Big)
  = 
  \chi+\chi^{2}+\chi^{5}+\chi^{7}+\chi^{10}+\chi^{11}
\end{multline}
is the $6$-dimensional representation of $\Z_{12}$ of cohomology
degree $2\equiv 0\modop 2$ generated by
\begin{equation}
  \diag\Big( 
  e^{\frac{2\pi i}{12}}, 
  e^{\frac{4\pi i}{12}}, 
  e^{\frac{10\pi i}{12}}, 
  e^{\frac{14\pi i}{12}}, 
  e^{\frac{20\pi i}{12}}, 
  e^{\frac{22\pi i}{12}}
  \Big)
  \,.
\end{equation}
In the future, we are just going to write $\otimes$, and it will be
understood that we are tensoring over $\C$.

\subsection{Minimal Model as Coset}
\label{sec:coset}

The minimal models for the ${\cal N}=2$ superconformal algebra have
equivalent realizations in terms of super-GKO coset models  
\begin{equation}
  \label{eq:GKO}
  \frac{
    \sufrak(2)_k \oplus \ufrak(1)_2 
  }{
    \ufrak(1)_{k+2}
  }
  \,,
\end{equation}
as well as Landau-Ginzburg models.  The modular invariant partition
functions fall into an ADE classification~\cite{Gepner:1987qi,
  GepnerQiu, Qiu}. From the coset CFT point of view these are obtained
from the ADE modular invariants of the $\sufrak(2)_k$ WZW model.  We
shall focus on the A series minimal models. There are various
subtleties concerning which modular invariant corresponds to the
A-type superpotential and it will turn out that there are essentially
four distinct models that will to of interest. The fields of the coset
CFT are labeled by $(j, n, s)$, where $j=0,\cdots, k/2$ is the
$\sufrak(2)_k$ highest weight, $n\in \Z_{2(k+2)}$ labels the
representations of the denominator $\ufrak(1)_{k+2}$ and $s\in \Z_4$
labels the free fermion representations in $\ufrak(1)_2$. There is a
$\Z_{k+2}\times \Z_2$ discrete group acting on the fields in the
following fashion
\begin{equation}
  \label{eq:action}
  \begin{split}
    \alpha:&~ \Phi_{(j, n, s)} \mapsto 
    (-1)^{\frac{2 n}{k+2}}\, \Phi_{(j,n,s)} 
    \,,
    \\
    \beta:&~ \Phi_{(j, n, s)} \mapsto 
    (-1)^{s} \, \Phi_{(j,n,s)} 
    \,.    
  \end{split}
\end{equation}
The $\Z_{k+2}$ action is realized geometrically in the gauged WZW
model by the rotation of the disc target space.  Orbifolding the
A-type theory with respect to these symmetries yields new modular
invariants, as was first observed in~\cite{MMSI}. Note that a
related issue arose in the context of WZW models for non-simply laced
groups in~\cite{Braun:2004qg}, where non-trivial automorphisms acting
on the fermions gave rise to new modular invariants for the
supersymmetric WZW models.

Since $s=1,3$ corresponds to the Ramond sector, the orbifold by
$\Z_2=\left<\beta\right>$ is from a space-time point of view the same
as modding out $(-1)^F$. The state space of the (charge conjugate)
diagonal modular invariant is 
\begin{equation}
  \Hilb_{MM_k} = 
  \bigoplus_{(j, n, s)} 
  \Hilb_{(j,n,s)} \otimes \overline{\Hilb}_{(j, n, s)} 
  \,,
\end{equation}
where the direct sum is over the standard range of super-parafermion
representations including the selection and identification rules
\begin{equation}
  \label{eq:fieldids}
  (j, n, s) \equiv 
  (k/2-j, n+k+2, s+2)
  \,,\qquad 
  2j+n+s \in 2\Z
  \,. 
\end{equation}
The state space of the $\Z_2$ orbifold is then
obtained as
\begin{equation}
  \Hilb_{MM_k/\Z_2} = 
  \bigoplus_{(j, n, s)}
  \Hilb_{(j,n,s)} \otimes \overline{\Hilb}_{(j, n, -s)} 
  \,.  
\end{equation}
Orbifolding $MM_k$ by $\Z_{k+2}\times \Z_2$ it was observed
in~\cite{MMSI} that the partition function is the same as in $MM_k$,
and that this model is in fact T-dual to $MM_k$. Likewise,
$MM_k/\Z_{k+2}$ is T-dual to $MM_k/\Z_2$.


Gepner models are orbifolds of tensor products of minimal models with
not necessarily equal level, which give rise to consistent,
GSO-projected string theory backgrounds. Consider a tensor product
of $r$ minimal models, of level $(k_1, \cdots k_r)$, and define
\begin{equation}
  \label{eq:GepnerDefs}
  \lambda = (j_1, \cdots, j_r) \,, \qquad
  \mu     = (n_1, \cdots, n_r;\,  s_1\, \cdots,s_r) \,,  
\end{equation}
and $\beta_{j}= (0,\cdots, 0,2,0 \cdots,0)$, with the non-zero entry at slot
$s_j$ and $\beta_0= (1 \cdots 1)$. Further define $K = \lcm (2, k_j + 2)$.
Then the partition function for the
Gepner model is given by
\begin{equation}
  \label{eq:GepnerPartitionFunction}
  Z_{(k_1, \cdots, k_r)} = 
  \sum_{\lambda, \mu} 
  \sum_{b_0= 0}^{2 K -1} 
  \sum_{b_j=0, 1} 
  \delta_\beta (-1)^{b_0} \chi_{\lambda, \mu} 
  \bar{\chi}_{\lambda, \mu + b_0 \beta_0 +\sum b_j \beta_j} 
  \,.
\end{equation}
The characters of the tensor product of the minimal models are denoted
by $\chi$. In principle, one can define the conserved D-brane charges
using RG flow~\cite{Fredenhagen:2002qn, MooreTalk}, but in practice
this is not feasible.

\subsection{Chern Character of the Minimal Model}
\label{sec:chern}

Now that we have defined all the ingredients, we can start to compute
the relevant K-groups. Our main tool is going to be the twisted
equivariant Chern character~\cite{FHTcomplex, MR2079380}. For
explicitness, let us consider a single minimal model whose
complexified D-brane charge group is
\begin{equation}
  \tK[\kappa]_{U(1)}\Big(SU(2)\Big)\otimes_\Z \C 
  ~\eqdef~
  \tK[\kappa]_{U(1)}\Big(SU(2); \C\Big)
  \,,
\end{equation}
where $\kappa=k+2$ is going to be the twist class\footnote{That is,
  the twist class is $\kappa$ times the generator of
  $H^3_{U(1)}\big(SU(2);\Z\big)$.} for the remainder of this section.
Now, given a twisted equivariant vector bundle we can tensor it with
any group representation, and get another equivariant vector bundle
with the same twist. In other words, there is an action of
$K_{U(1)}\big(\ptset;\C\big)=\R{U(1)}=\C[z,z^{-1}]$ on the twisted
equivariant K-theory.

In geometrical terms, $\C[z,z^{-1}]$ is the ring of functions on
$\C^\times\eqdef \C\setminus\{0\}$. And the twisted equivariant
K-theory $\tK[\kappa]_{U(1)}\big(SU(2); \C\big)$ is a module over this
function algebra, that is a sheaf over the base space $\C^\times$. The
twisted equivariant Chern character identifies the stalks (fibers) of
this sheaf over a point in $\C^\times$ with a certain cohomology
group.  More precisely, Freed-Hopkins-Teleman~\cite{FHTcomplex}
identify the stalk over $\zeta\in \C^\times$ with
\begin{equation}
  \label{eq:chMM}
  \tK[\kappa]_{U(1)}^\ast\Big(SU(2); \C\Big) \bigg|_\zeta
  ~\simeq~
  \tH[\kappa]_{U(1)}^\ast\Big(SU(2)^\zeta; 
  \tL[\kappa](\zeta)\Big)
  \,,
\end{equation}
where $\tL[\kappa](\zeta)$ is a certain flat line bundle. Note that
when we say that $\zeta$ acts on $SU(2)$, we really mean that
$\frac{\zeta}{|\zeta|}\in U(1)$ acts on $SU(2)$.

In general the knowledge of the stalks is not enough to reconstruct
the sheaf, for example every fiber of a line bundle is just isomorphic
to $\C$. However, in the case of a single minimal model the sheaf
turns out to be a skyscraper sheaf, and can indeed be reconstructed.

\subsection{Twisted Equivariant Cohomology of the Minimal Model}
\label{sec:cohomology}

In this section, we are going to determine the twisted equivariant
cohomology groups that appear in the Chern character formula
eq.~\eqref{eq:chMM}.  We advise the reader who is not interested in
all the details to note the result, eqns.~\eqref{eq:twistedcohMMA}
and~\eqref{eq:twistedcohMMB}, and then proceed with the next section.

In fact, the problem is very similar to
$\tK[\kappa]_{SU(2)}\big(SU(2); \C\big)$ which is explicitly worked
out as an example in~\cite{FHTcomplex}. Depending on $\zeta$, there
are two different fixed point sets. One possibility is $\zeta\in
\IR_{>0}$, which acts trivially on the whole $SU(2)$. It turns
out~\cite{FHTcomplex} that the line bundle $\tL[\kappa](\zeta)$ is
trivial in that case. Therefore, the \emph{untwisted} equivariant
cohomology is
\begin{equation}
  H_{U(1)}^\ast\Big(SU(2)^\zeta;
  \tL[\kappa](\zeta)\Big)
  = 
  H_{U(1)}^\ast
  \Big(SU(2); \C \Big)
  = 
  \C[u,t]\big/u^2
  \,,
\end{equation}
where we used the Leray spectral sequence 
\begin{equation}
  H^p\Big( BU(1); H^q\big(SU(2);\C\big) \Big) 
  ~\Rightarrow~
  H^{p+q}_{U(1)}\big(SU(2);\C\big)
\end{equation}
with $t\in H^2\big(BU(1);\C\big)$ of degree $2$ and $u\in
H^3\big(SU(2);\C\big)$ of degree $3$. To determine the twisted
equivariant cohomology $\tH[\kappa]_{U(1)}^\ast \big(SU(2); \C \big)$
from the untwisted one, we have to mod out\footnote{More formally, we
  are using the untwisted to twisted cohomology spectral sequence.} by
the additional differential\footnote{Note that $(d_3)^2\sim u^2= 0$ in
  $\C[u,t]/u^2$.} $d_3 = \kappa u$. An easy computation shows that 
\begin{equation}
  \tH[\kappa]_{U(1)}^\ast\Big(SU(2)^\zeta;
  \tL[\kappa](\zeta)\Big)
  = 
  \tH[\kappa]_{U(1)}^\ast
  \Big(SU(2); \C \Big)
  =
  \frac{\ker\big( d_3 \big)}{\img\big( d_3 \big)} 
  = 
  0
  \,.
\end{equation}
This settles the case where the whole $SU(2)$ is fixed under the
$\zeta$ action. The other possibility is the generic case where
$SU(2)^\zeta= \SA$. In that case, the flat line bundle
$\tL[\kappa](\zeta)$ over $\SA$ has~\cite{FHTcomplex} holonomy
$\zeta^\kappa$, so all cohomology groups vanish unless
$\zeta^\kappa=1$. In that case, that is over the $\kappa-1$ points
\begin{equation}
  \zeta_m
  \eqdef
  e^{\frac{2\pi i m}{\kappa}}
  \,,\quad
  m=1,\dots, \kappa-1 
\end{equation}
the untwisted cohomology is
\begin{multline}
  \label{eq:HSAzetam}
  H_{U(1)}^\ast\Big(\SA;
  \tL[\kappa](\zeta_m)\Big)
  = 
  H^\ast\Big(BU(1);\C\Big)
  \otimes 
  H^\ast\Big(\SA; \tL[\kappa](\zeta_m) \Big)
  = \\ =
  \C[t]\otimes
  \C[v]\big/v^2
  =
  \C[v,t]\big/v^2
  \,,
\end{multline}
where $\deg(v)=1$ and $\deg(t)=2$. The twist class is in
$H_{U(1)}^3\big(\SA; \tL[\kappa](\zeta_m)\big) = \C\cdot tv$. Hence, 
if one normalizes $tv$ properly then $d_3=\kappa tv$. The
$d_3$-cohomology is
\begin{equation}
  \tH[\kappa]_{U(1)}^\ast\Big(\SA;
  \tL[\kappa](\zeta_m)\Big)
  = 
  \frac{\ker\big( d_3 \big)}{\img\big( d_3 \big)} 
  = 
  \C v
  = 
  \begin{cases}
    \C
    \,,& 
    \ast = 1
    \,; \\
    0
    \,, & 
    \text{else.}
  \end{cases}
\end{equation}

In addition to the $U(1)$ action on $SU(2)$, we can also act with
$\Z_\kappa$. We find two more cases, the fixed point set can be either
$\SB$ or empty. The cohomology of the empty set of course vanishes. In
the former case, note that $U(1)$ acts simply transitive on $\SB$, so
the equivariant cohomology is just the cohomology of a point. To
summarize, there are four different cases corresponding to different
$g\in U(1)\times \Z_\kappa$. The twisted cohomology groups are
(ignoring the $\Z_\kappa$ action on the cohomology and the precise
degrees for now)
\begin{subequations}
  \begin{align}
    g =&~
    (1,0)
    & 
    \Rightarrow~
    \tH[\kappa]_{U(1)}^\ast\big(SU(2)^g;\tL[\kappa](g)\big) =&~
    \tH[\kappa]_{U(1)}^\ast\big(SU(2);\C\big) = 0
    \,,
    \\
    \label{eq:gzeta0}
    g =&~
    (\zeta,0)
    & 
    \Rightarrow~
    \tH[\kappa]_{U(1)}^\ast\big(SU(2)^g;\tL[\kappa](g)\big) =&~
    \tH[\kappa]_{U(1)}^\ast\big(\SA;\tL[\kappa](g)\big) = 
    \delta_{\zeta^\kappa, 1} \C
    \,,
    \\
    g =&~ (1,n)
    & 
    \Rightarrow~
    \tH[\kappa]_{U(1)}^\ast\big(SU(2)^g;\tL[\kappa](g)\big) =&~
    H_{U(1)}^\ast\big(\SB;\C\big) = 
    H^\ast\big(\ptset\big) = \C
    \,,
    \\
    g =&~ (\zeta,n)
    & 
    \Rightarrow~
    \tH[\kappa]_{U(1)}^\ast\big(SU(2)^g;\tL[\kappa](g)\big) =&~
    H_{U(1)}^\ast\big(\varnothing;\C\big) = 
    0
    \,,
  \end{align}
\end{subequations}
where we took $n\in \Z_\kappa\setminus\{0\}$ and $\zeta \in
\C^\times\setminus\{1\}$. 

All that remains is to determine the precise action of $\Z_\kappa$ on
the cohomology group eq.~\eqref{eq:gzeta0}. For that, note that even
though the line bundle $\tL[\kappa](\zeta_m)$ in
eq.~\eqref{eq:HSAzetam} is trivial, the trivializing section winds $m$
times around the $\SA$ relative to \emph{the} trivial line bundle.
Therefore rotating $\SA$ by $\frac{2\pi}{\kappa}$ multiplies $v$ with
the phase $\exp(\frac{2\pi i m}{\kappa})$. In terms of the character
$\chi:\Z_\kappa\to U(1),~m\mapsto \exp(\frac{2\pi i m}{\kappa})$ this
means that
\begin{equation}
  \bigoplus_{\zeta\in\C^\times}
  \tH[\kappa]_{U(1)}^\ast\Big(SU(2)^\zeta;\tL[\kappa](\zeta)\Big) =
  \left\{
    \begin{array}{cl}
      0
      \,, &
      \ast = 0
      \\
      \chi + \chi^2 + \cdots + \chi^{\kappa-1}
      \,, &
      \ast = 1
    \end{array}
  \right\}
  = 
  \RZt_\kappa
\end{equation}
as $\Z_\kappa$ representation. In other words, we can write the
twisted equivariant cohomology groups as
\begin{subequations}
  \begin{align}
    \label{eq:twistedcohMMA}
    n = 0     ~&\in \Z_\kappa 
    & 
    \Rightarrow&~
    \bigoplus_{\zeta\in\C^\times}
    \tH[\kappa]_{U(1)}^\ast\Big(SU(2)^{(\zeta,n)};
    \tL[\kappa](\zeta,n)\Big) 
    =    
    \RZt_\kappa
    \\
    \label{eq:twistedcohMMB}
    n \not= 0 ~&\in \Z_\kappa    
    &
    \Rightarrow&~
    \bigoplus_{\zeta\in\C^\times}
    \tH[\kappa]_{U(1)}^\ast\Big(SU(2)^{(\zeta,n)};
    \tL[\kappa](\zeta,n)\Big) 
    =    
    \C
  \end{align}
\end{subequations}
using the conventions for cohomology degrees in
eqns.~\eqref{eq:degC},~\eqref{eq:degRZ}.

\subsection{Mirror Symmetry for Minimal Models}
\label{sec:minimirror}

As a quick application, let us compute the K-groups of the minimal
model and its $\Z_\kappa$ orbifold. According to the twisted
equivariant Chern character, the K-groups of the minimal model are
\begin{equation}
  \label{eq:MMA}
  \tK[\kappa]_{U(1)}^\ast\Big( SU(2); \C\Big) 
  =
  \bigoplus_{\zeta \in \C^\times}  
  \tH[\kappa]_{U(1)}^\ast\Big( SU(2)^\zeta; \tL[\kappa](\zeta)\Big)
  = 
  \RZt_\kappa
  \simeq 
  \begin{cases}
    0
    \,, &
    \ast = 0
    \,; \\
    \C^{\kappa-1}
    \,, &
    \ast = 1
    \,,
  \end{cases}
\end{equation}
using the cohomology groups computed in eq.~\eqref{eq:twistedcohMMA}.
We recover the known~\cite{Schafer-Nameki:2003aj} D-brane charge
groups for the coset minimal model.

Similarly, we can compute the D-brane charge group in the $\Z_\kappa$
orbifold which is known to be the mirror of the minimal model. One
obtains
\begin{equation}
  \label{eq:MMB}
  \begin{split}
    \tK[\kappa]_{U(1)\times\Z_\kappa}^\ast\Big( SU(2); \C\Big) 
    =&~
    \bigoplus_{(\zeta,n) \in \C^\times\times\Z_\kappa}  
    \tH[\kappa]_{U(1)\times\Z_\kappa}^\ast\Big( SU(2)^{(\zeta,n)}; 
    \tL[\kappa](\zeta,n)\Big)
    = \\ =&~
    \bigoplus_{n \in \Z_\kappa}  
    \bigoplus_{\zeta \in \C^\times}  
    \tH[\kappa]_{U(1)\times\Z_\kappa}^\ast\Big( SU(2)^{(\zeta,n)}; 
    \tL[\kappa](\zeta,n)\Big)
    = \\ =&~
    \bigoplus_{n \in \Z_\kappa}  
    \left[
      \bigoplus_{\zeta \in \C^\times}  
      \tH[\kappa]_{U(1)}^\ast\Big( SU(2)^{(\zeta,n)}; 
      \tL[\kappa](\zeta,n)\Big)
    \right]^{\Z_\kappa}
    = \\ =&~
    \left[
    \underbrace{\RZt_\kappa}_{n=0}
    \oplus 
    \underbrace{\C}_{n=1}
    \oplus 
    \cdots
    \oplus 
    \underbrace{\C}_{n=\kappa-1}
    \right]^{\Z_\kappa}
    = \\ =&~
    \C^{\kappa-1}
    \simeq 
    \begin{cases}
      \C^{\kappa-1}
      \,, &
      \ast = 0
      \,; \\
      0
      \,, &
      \ast = 1
      \,.
    \end{cases}    
  \end{split}
\end{equation}
Note that the $\Z_\kappa$ equivariant cohomology is simply the
$\Z_\kappa$ invariant subspace of the cohomology group. For that, it
is important to work with complex coefficients, because it would
generate torsion contributions over the integers. Also note that
the $\Z_\kappa$ equivariant K-theory is in general \emph{not} the same
as the $\Z_\kappa$ invariant K-groups. 

To summarize, we observe that the $\Z_\kappa$ orbifold indeed
exchanges $K^0\leftrightarrow K^1$, as we expect from the mirror
involution. Furthermore, recall the distinction between A- and B-type
branes~\cite{MMSI}. The A-branes carry the charges in
eq.~\eqref{eq:twistedcohMMA}, contributing to $K^1$ of the minimal
model. On the other hand side, the B-branes
eq.~\eqref{eq:twistedcohMMB} are only stable in the $\Z_\kappa$
orbifold of the minimal model where they contribute to $K^0$.

\subsection{K-Groups for Gepner Models}
\label{sec:kgroups}

Having tackled a single minimal model, we now proceed to Gepner
models~\cite{Gepner:1987qi, GepnerQiu, Qiu, Recknagel:1997sb}. For
that we take $d$ copies of the $SU(2)$ with the action of $d$ copies
of $U(1)$ factor-by-factor. That is
\begin{equation}
  U(1)^d \times SU(2)^d \to SU(2)^d
\end{equation}
with a choice of twist
\begin{equation}
  \overline{\kappa} = (\kappa_1,\kappa_2,\dots, \kappa_d)
  \,,
\end{equation}
where $k_i=\kappa_i-2$ is the level in the CFT of the $i$-th
factor. The overall central charge is 
\begin{equation}
  c = 
  \sum_{i=1}^d
  \frac{3 k_i}{k_i+2} 
  =
  \sum_{i=1}^d
  \frac{3 (\kappa_i-2)}{\kappa_i} 
  \,.
\end{equation}
Whenever $\frac{c}{3}$ is integer, this could be the central charge of
a geometric compactification of that dimension. However, a mere tensor
product of minimal models is never geometric because of non-integer
charges. In other words, it does not have space-times
supersymmetry. The solution to this problem~\cite{Gepner:1987qi} is to
orbifold by a certain discrete symmetry group $\GSO$. 

As we have seen, each of the minimal models has a discrete symmetry
group $\Z_{\kappa_i}=\{0,1,\dots,\kappa_i-1\}$. The GSO projection is
the group generated by 
\begin{equation}
  \big(
  1,1,\dots,1
  \big)
  \in 
  \prod_{i=1}^d \Z_{\kappa_i}
  \,.
\end{equation}
It follows that
\begin{equation}
  \label{eq:GSO}
  \GSO = \Z_{ \lcm(\kappa_1,\kappa_2,\dots, \kappa_d)}
  \,.
\end{equation}
According to the general dictionary between D-brane charge groups and
K-theory group, the D-brane charges in the Gepner model are
\begin{equation}
  \tK[\overline{\kappa}]_{U(1)^d\times \GSO}\Big( SU(2)^d \Big)
  \,.
\end{equation}
We can again compute (the complexification) through the twisted
equivariant Chern character. Once we translate the K-groups into
cohomology, we can use that
\begin{itemize}
\item the $\GSO$ equivariant cohomology is the $\GSO$ invariant
  cohomology and
\item the K\"unneth theorem for cohomology,
\end{itemize}
neither of which hold in general for twisted equivariant
K-theory. Again, we have to complexify
\begin{equation}
  U(1)^d \times \GSO 
  \quad\rightsquigarrow\quad
  \big(\C^\times\big)^d \times \GSO
\end{equation}
and think of the cohomology and K-groups as sheaves over this space.
According to Section~\ref{sec:cohomology}, the only potentially
non-vanishing cohomology groups for the $i$-th minimal model sit over
the $\kappa_i$-th roots of unity
\begin{equation}
  \mathcal{Z}_i
  \eqdef
  \Big\{
  e^{\frac{2\pi i m}{\kappa_i}}
  \Big|
  m\in \Z_{\kappa_i}=\{0,\dots,\kappa_i-1\}
  \Big\}
  \subset \C^\times  
  \,,
\end{equation}
therefore the only non-vanishing cohomology groups of the product are
over the points 
\begin{equation}
  \mathcal{Z}
  \eqdef
  \prod_{i=1}^d 
  \mathcal{Z}_i
  =
  \Big\{
  \big( e^{\frac{2\pi i m_1}{\kappa_1}}, \dots, 
  e^{\frac{2\pi i m_d}{\kappa_d}}
  \Big|
  m_i \in \Z_{\kappa_i}
  \Big\}
  \subset \big(\C^\times\big)^d
  \,.
\end{equation}
Using all that we obtain
\begin{multline}
  \label{eq:Kgepner1}
  \tK[\overline{\kappa}]^\ast_{U(1)^d}\Big(SU(2)^d; \C\Big)
  =
  \bigoplus_{g\in \GSO}
  \left[
    \bigoplus_{\overline{z}\in (\C^\times)^d}
    \tH[\overline{\kappa}]_{U(1)^d}^\ast
    \left(
      \bigtimes_{i=1}^d
      SU(2)^{(z_i,g)}
      ;
      \bigotimes_{i=1}^d
      \tL[\kappa_i](z_i,g)
    \right)  
  \right]^\GSO
  = \\ =
  \bigoplus_{g\in \GSO}
  \left[
    \bigoplus_{\overline{z}\in \mathcal{Z}}
    \tH[\overline{\kappa}]_{U(1)^d}^\ast
    \left(
      \bigtimes_{i=1}^d
      SU(2)^{(z_i,g)}
      ;
      \bigotimes_{i=1}^d
      \tL[\kappa_i](z_i,g)
    \right)  
  \right]^\GSO
  = \\ =
  \bigoplus_{g\in \GSO}
  \left[
    \bigotimes_{i=1}^d 
    \left\{
      \bigoplus_{z_i\in \mathcal{Z}_i}
      \tH[\kappa_i]_{U(1)}^\ast
      \Big(
      SU(2)^{(z_i,g)}
      ;
      \tL[\kappa_i](z_i,g)
      \Big)
    \right\}
  \right]^\GSO
\end{multline}
Note that according to eqns.~\eqref{eq:twistedcohMMA}
and~\eqref{eq:twistedcohMMB}, 
\begin{equation}
  \bigoplus_{z_i\in \mathcal{Z}_i}
  \tH[\kappa_i]_{U(1)}^\ast
  \Big(
  SU(2)^{(z_i,g)}
  ;
  \tL[\kappa_i](z_i,g)
  \Big)
  = 
  \begin{cases}
    \RZt_{\kappa_i}
    \,, & 
    g \equiv 0 \modop \kappa_i 
    ~\Leftrightarrow~
    \kappa_i \mid g
    \,;\\
    \C
    \,, & 
    \kappa_i \nmid g
    \,.
  \end{cases}
\end{equation}
Moreover, $\GSO$ obviously acts on $\RZt_{\kappa_i}$ as
$\p{\Z_{\kappa_i}}{\GSO}\big( \RZt_{\kappa_i} \big)$, see
eq.~\eqref{eq:pdef}. Therefore, we can simplify
eq.~\eqref{eq:Kgepner1} to
\begin{equation}
  \label{eq:Kgepner}
  \tK[\overline{\kappa}]^\ast_{U(1)^d}\Big(SU(2)^d; \C\Big)
  =
  \bigoplus_{g\in \GSO}
  \left[
    \bigotimes_{\kappa_i \mid g}
    \p{\Z_{\kappa_i}}{\GSO}\Big( \RZt_{\kappa_i} \Big)
  \right]^\GSO
  \,,
\end{equation}
where we would like to remind the reader that $n\mid 0$ for all $n$,
that $\otimes_{i\in\varnothing}=\C$, and that we defined $\RZt_{\kappa_i}$ to
have cohomological degree $1$.

\section{Examples}
\label{sec:examples}

\subsection{Toroidal Theories}
\label{sec:torus}

There are three Gepner models~\cite{WendlandThesis} that describe an
elliptic curve. Two of them, $k=(1,1,1)$ and $k=(0,1,4)$, turn out to
be the same CFT (for example, have identical partition functions).
Hence, we obtain two different CFTs corresponding to the two orbifold
singularities in the complex structure moduli space of the torus, see
Table~\ref{tab:ellipticcurves}. Recall that each elliptic curve
$\C/(\Z\oplus\tau \Z)$ has a $\Z_2$ symmetry, but at $\tau=i$ and
$\tau=\exp({\frac{2\pi i}{3}})$ the symmetry is enhanced to $\Z_4$ and
$\Z_6$, respectively.
\begin{table}[htbp]
  \centering
  \begin{tabular}{cccc}
    Complex structure &    
    Symmetry & 
    Gepner model & 
    Hypersurface
    \\ \hline
    $\tau = i$ & $\Z_4$ & $k=(0,2,2)$ &
    $\{x_0^2+x_1^4+x_2^4=0\} \subset \WP_{2,1,1}$ \\
    $\tau = e^{\frac{2\pi i}{3}}$ & $\Z_6$ & $k=(1,1,1)$ &
    $\{x_0^3+x_1^3+x_2^3=0\} \subset \WP_{1,1,1}$ \\
    $\tau = e^{\frac{2\pi i}{3}}$ & $\Z_6$ & $k=(0,1,4)$ &
    $\{x_0^2+x_1^3+x_2^6=0\} \subset \WP_{3,2,1}$ \\
  \end{tabular}
  \caption{Elliptic curves with enhanced automorphism groups.}
  \label{tab:ellipticcurves}
\end{table}
We easily compute using eq.~\eqref{eq:Kgepner} that in all three cases
\begin{equation}
  \tK[\overline{\kappa}]_{U(1)^3\times \GSO}^\ast\Big(SU(2)^3;\C\Big) = 
  \left\{
    \begin{array}{cc}
      \C^2 
      \,,& 
      \ast=0
      \\
      \C^2 
      \,,& 
      \ast=1
    \end{array}
  \right\}
  = 
  K^\ast\big(T^2;\C\big)
  \,,
\end{equation}
as expected since we are dealing with a topological invariant of the
torus. Note that the toroidal Gepner models have always $3$ factors,
even if that forces one of the levels to be zero. It is important to
realize~\cite{Brunner:2004zd} that adding one factor with $c=0$ in the
Gepner model does indeed have a physical effect. For example, we can
easily compute the D-brane charges in the $k=(2,2) \Leftrightarrow
\kappa=(4,4)$ model and obtain
\begin{equation}
  \tK[(4,4)]_{U(1)^2\times \GSO}^\ast\Big(SU(2)^2;\C\Big) = 
  \begin{cases}
    \C^6
    \,, & 
    \ast=0
    \,; \\
    0 
    \,, & 
    \ast=1
    \,.
  \end{cases}
\end{equation}
This is not the D-brane charge group of any geometric $c=3$ CFT. Note
that the usual argument why $k=0$ factors do not matter is wrong: In
the corresponding Landau-Ginzburg model~\cite{Brunner:2004mt}, the
$k=0$ factor corresponds to a field $\Phi$ which appears in the
superpotential as
\begin{equation}
  W_\text{LG} = \cdots + \Phi^2
  \,.
\end{equation}
Then it is claimed that one can integrate out $\Phi$ at no cost. But
that is only true if one restricts to the closed strings, if one
considers D-branes and open strings then one must include a boundary
action which will contain $\Phi$ as well.

\subsection{Twisted Sectors}
\label{sec:twisted}

Let us have a closer look at the formula for the K-groups of a Gepner
model, eq.~\eqref{eq:Kgepner}. First, let us rewrite it as
\begin{equation}
  \label{eq:KgepnerRepeat}
  \tK[\overline{\kappa}]^\ast_{U(1)^d}\Big(SU(2)^d; \C\Big)
  =
  \bigoplus_{g\in \GSO}
  \Big(\Kc_g\Big)^\GSO
\end{equation}
with 
\begin{equation}
  \Kc_g \,\eqdef~
  \bigotimes_{\kappa_i \mid g}\,
  \p{\Z_{\kappa_i}}{\GSO}\Big( \RZt_{\kappa_i} \Big)
  \,.
\end{equation}
Obviously, this has an interpretation of $\Kc_g^\GSO$ being the
contribution of the $g$-twisted sector in the $\GSO$ orbifold.  Note
that a single tensor factor $\p{\Z_{\kappa_i}}{\GSO}\big(
\RZt_{\kappa_i} \big)$ does not have any $\GSO$-invariant subspace, so
the only way to obtain something invariant is to either have zero
factors (which yields a B-type brane), or $\geq 2$ factors. This is
very familiar from the geometric interpretation as hypersurfaces in
weighted projective spaces. If two or more weights
$\frac{|\GSO|}{\kappa_i}$ have a common factor, then the Calabi-Yau
hypersurface inherits an orbifold singularity from the ambient space.
The exceptional divisor from the resolution of the singularity
increases the rank of the K-groups.

Specifically, in complex dimension $\geq 2$ one can have genuine
singularities which require resolutions and contribute twisted sector
D-brane charges. To see that explicitly within the Gepner model
context, let us consider the following two $K3$ Gepner models. First
consider the $(k=2)^4$ Gepner model, corresponding to the Fermat
quartic
\begin{equation}
  \big\{
  x_0^4 + x_1^4 + x_2^4 + x_3^4 =0
  \big\} \subset \CP^3
  \,.
\end{equation}
In this case, the ambient space and the hypersurface are
non-singular. The contribution of the untwisted and the three
$g$-twisted sectors is 
\begin{equation}
  \begin{array}{c||cccc}
    g \in \GSO & 0 & 1 & 2 & 3 
    \\[1ex] 
    \Kc_g& 
    \big( \RZt_{4} \big)^4 & \C & \C & \C
    \\[1ex]
    \dim_\C\Kc_g^\GSO & 
    21 & 1 & 1 & 1
    \\[1ex] 
    \text{Type} &
    A^4 & B^4 & B^4 & B^4 
  \end{array}
  \Rightarrow
  \lrstack[t]{
    \tK[(4,4,4,4)]^\ast_{U(1)^4}\Big(SU(2)^4; \C\Big) 
    = 
    }{=
      \begin{cases}
        \C^{24}
        \,, & 
        \ast=0
        \,; \\
        0 
        \,, & 
        \ast=1
        \,.
      \end{cases}  
    }
\end{equation}
We can do the same for the $k=(1,2,2,4) \Leftrightarrow
\kappa=(3,4,4,6)$ Gepner model. It corresponds to the singular $K3$
hypersurface
\begin{equation}
  X\eqdef
  \big\{ 
  x_0^3 + x_1^4 + x_2^4 + x_3^6 =0
  \big\} \subset \WP_{4,3,3,2}
\end{equation}
The weighted projective space has a rational curve $C_2$ of
$\C^2/\Z_2$ singularities and another rational curve $C_3$ of
$\C^2/\Z_3$ singularities embedded as
\begin{equation}
  \begin{split}
    C_2
    ~&\hookrightarrow
    \WP_{4,3,3,2}
    \,,~
    [s_0,s_1] \mapsto \big[ s_0, 0,0, s_1 \big]
    \,,
    \\
    C_3
    ~&\hookrightarrow
    \WP_{4,3,3,2}
    \,,~
    [s_0,s_1] \mapsto \big[ 0, s_0, s_1, 0 \big]
    \,.
  \end{split}  
\end{equation}
The surface inherits $4A_1$ and $6A_2$ orbifold singularities from
\begin{equation}
  C_2 \cap X = 4
  \,,\quad 
  C_3 \cap X = 6
  \,.
\end{equation}
The resolution $\widetilde{X}$ is then a smooth $K3$ surface. This
concludes the geometric point of view, now let us analyze the K-theory
computation from the Gepner model side. Using
eq.~\eqref{eq:KgepnerRepeat}, we find
\begin{equation}
  \begin{array}{c||cccccccccccc}
    g \in \GSO & 
    0 & 1 & 2 & 3 & 4 & 5 & 6 & 7 & 8 & 9 & 10 & 11
    \\[1ex] 
    \Kc_g & 
    \RZt_{3,4,4,6} &
    \C & 
    \C & 
    \RZt_{3} &
    \RZt_{4,4} &
    \C & 
    \RZt_{3,6} &
    \C & 
    \RZt_{4,4} &    
    \RZt_{3} &
    \C & 
    \C     
    \\[1ex]
    \dim_\C\Kc_g^\GSO & 
    10 & 1 & 1 & 0 & 3 & 1 & 2 &1 & 3 & 0 & 1 & 1
    \\[1ex] 
    \text{Type} &
    A^4 & B^4 & B^4 & - & 
    {\scriptstyle BA^2B} & B^4 & 
    {\scriptstyle AB^2\!A} & B^4 & 
    {\scriptstyle BA^2B} & - & B^4 & B^4 
  \end{array}  
\end{equation}
where we abbreviated
\begin{equation}
  \RZt_{\kappa_1,\kappa_2,\dots} 
  \eqdef 
  \bigotimes_{i=1,2,\dots}
  \p{\Z_{\kappa_i}}{\GSO}\Big( \RZt_{\kappa_i} \Big)  
  \,.
\end{equation}
Of course, in the end we obtain again the K-groups of the $K3$
manifold.  However, this times some of the charge groups involve
mixtures of A- and B-type branes. In the same way one can analyze all
$K3$ Gepner models, see Appendix~\ref{sec:tableK3}.

\section{Kn\"orrer Periodicity}
\label{sec:knorrer}

If one adds two variables with a quadratic superpotential to the
Landau-Ginzburg theory~\cite{Witten:1993yc, Orlov, HoriRG} with fields
$\Phi=(\phi_1,\dots)$,
\begin{equation}
  W_\text{LG}\big(\Phi\big) 
  \longrightarrow
  \widehat{W}_\text{LG} =
  W_\text{LG}\big(\Phi\big) 
  + x^2 +y^2
  \,,
\end{equation}
then one obtains the same theory again. This is quite non-trivial,
because adding a single variable with a quadratic superpotential
certainly does yield an inequivalent theory as discussed in
Section~\ref{sec:torus}.

The evidence for periodicity is that the topological B-branes, that is
the category of matrix factorizations, are equivalent. This fact is
known as Kn\"orrer periodicity~\cite{MR877010},
\begin{equation}
  \mathbf{MF}\Big( \C[\Phi]\big/W_\text{LG}(\Phi) \Big)
  \simeq
  \mathbf{MF}\Big( 
  \C[\Phi,x^2,y^2]\big/\widehat{W}_\text{LG}(\Phi,x,y) \Big)
  \,.
\end{equation}
This periodicity manifests itself in our formula
eq.~\eqref{eq:Kgepner} as follows. Adding two factors with
$k=0\Leftrightarrow \kappa=2$ amounts to inserting
\begin{equation}
  \p{\Z_{2}}{\GSO}\Big( \RZt_{2} \Big)    
  \otimes
  \p{\Z_{2}}{\GSO}\Big( \RZt_{2} \Big)    
  = 
  \C
\end{equation}
whenever $2\mid g$. But 
\begin{equation}
  \big(\cdots\big) \otimes \C = \big(\cdots\big)
\end{equation}
is the identity, so we obtain again the same K-groups. 

Note that the above argument is flawed since adding the $\kappa=2$
factors might change the $\GSO$ group eq.~\eqref{eq:GSO}. If the
initial order $|\GSO|$ was odd, that is,
\begin{equation}
  \lcm(\kappa_1,\dots,\kappa_d) \in 2\Z+1
  \,,
\end{equation}
then 
\begin{equation}
  \lcm(\kappa_1,\dots,\kappa_d,2,2) =
  2 \lcm(\kappa_1,\dots,\kappa_d)
  \,.
\end{equation}
Therefore, periodicity only holds if one had already an even
$\kappa_i$. In general, Kn\"orrer periodicity need not hold for the
first time one adds two $k=0$ factors, but it always holds from the
second time onward,
\begin{equation}
  \tK[(\kappa_1,\dots,\kappa_d,2)]_{U(1)^{d+1}\times \GSO}
  \Big( SU(2)^{d+1} ; \C\Big) 
  = 
  \tK[(\kappa_1,\dots,\kappa_d,2,2,2)]_{U(1)^{d+3}\times \GSO}
  \Big( SU(2)^{d+3} ; \C\Big) 
  \,.
\end{equation}
This is somewhat reminiscent of stabilization in K-theory.

\section{Generalized Permutation Branes}
\label{sec:mrg}

In this section, we are going to focus on the Calabi-Yau ($c=9$)
Gepner models.  It is clear from Section~\ref{sec:minimirror} that all
D-brane charges can be found as suitable combinations of the D-branes
in the coset or its mirror ($\Z_\kappa$ orbifold). In particular, the
usual tensor product and permutation branes give all the D-brane
charges in the untwisted sector, corresponding to $g=0$ in
eq.~\eqref{eq:Kgepner}.  Similarly, one obtains zero or one brane in
the twisted ($g=1,\dots,|\GSO|$) sectors. But the latter is not enough
to fill out the D-brane charge lattice in general, since sometimes
there are two or more independent charges coming from a twisted
sector. Of course, all that means is that the boundary state
construction is incomplete. Using Landau-Ginzburg models and matrix
factorizations one obtains~\cite{Walcher:2004tx, Caviezel:2005th} all
brane charges.

Inspection of the formula for the K-groups eq.~\eqref{eq:Kgepner}
shows that $2$ or more brane charges can only come from a $g\in \GSO$
sector where some $\kappa_i$ divides $g$. Moreover, if only a single
$\kappa_i$ divides $g$, then there is no contribution because
\begin{equation}
  \left[
  \p{\Z_{\kappa_i}}{\GSO}\Big({\RZt_{\kappa_i}}\Big)
  \right]^\GSO 
  = 0
\end{equation}
has no invariant subspace. Hence, the interesting case is if two or
more $\kappa_i$ have a common factor. Following~\cite{Caviezel:2005th}
let us consider the case where $r$ of the shifted levels
$\overline{\kappa}=(\kappa_1,\dots)$ have the same\footnote{If the
  common divisor $d=2$, then there is again only a one-dimensional
  contribution to the K-group in the $g\in d\Z$ twisted sectors, which
  is not so interesting. Of course, our arguments hold in that case as
  well.} divisor $d>2$.

First, note that $r$ odd contributes to $K^1$ only, as is evident from
our degree convention eqns.~\eqref{eq:degC},~\eqref{eq:degRZ}. Not so
surprisingly, if one~\cite{Caviezel:2005th} restricts oneself to $K^0$
then there are no D-brane charges for $r=1,3,5$. This leaves the cases
$r=2$ and $r=4$. Looking at the list of Gepner models, $r=4$ can only
occur if the Gepner model has more than $5$ minimal model factors.
There is nothing wrong with that, and our formula
eq.~\eqref{eq:Kgepner} gives the correct answer for the K-groups.
However, if one~\cite{Caviezel:2005th} were to restrict oneself to $5$
minimal model factors, then $r=4$ cannot occur either.

\section{Conclusions}
\label{sec:conclusions}

There is a very simple formula eq.~\eqref{eq:Kgepner} for the rank of
the K-groups of Gepner models. The summands in the formula have a
natural interpretation as the contributions from twisted sectors. We
checked the computation in $c=3,6,9$ Gepner models and find agreement
with the topology of the associated Calabi-Yau manifolds.

\appendix

\section{K3 Gepner Models}
\label{sec:tableK3}

There are $16$ Gepner models which are associated to $K3$
surfaces~\cite{MR1810775, MR1950953, MR2029366} listed in
Table~\ref{tab:K3}.
\begin{table}[htbp]
  \centering
  \begin{tabular}{llll}
    $\overline{k}=(1,1,1,1,1,1)$, & 
    $\overline{k}=(0,1,1,1,1,4)$, &
    $\overline{k}=(2,2,2,2)$,   &
    $\overline{k}=(1,2,2,4)$,   \\
    $\overline{k}=(1,1,4,4)$,   &
    $\overline{k}=(1,1,2,10)$,  &
    $\overline{k}=(0,4,4,4)$,   &
    $\overline{k}=(0,3,3,8)$,   \\
    $\overline{k}=(0,2,6,6)$,   &
    $\overline{k}=(0,2,4,10)$,  &
    $\overline{k}=(0,2,3,18)$,  &
    $\overline{k}=(0,1,10,10)$, \\
    $\overline{k}=(0,1,8,13)$,  &
    $\overline{k}=(0,1,7,16)$,  &
    $\overline{k}=(0,1,6,22)$,  &
    $\overline{k}=(0,1,5,40)$
  \end{tabular}
  \caption{Gepner models associated to $K3$.}
  \label{tab:K3}
\end{table}
We checked that we obtain
\begin{equation}
  \tK[\overline{\kappa}]_{U(1)^d\times \GSO}^\ast\Big(SU(2)^d;\C\Big) = 
  \left\{
    \begin{array}{cc}
      \C^{24}
      \,,& 
      \ast=0
      \\
      0
      \,,& 
      \ast=1
    \end{array}
  \right\}
  = 
  K^\ast\big(K3;\C\big)
  \,.  
\end{equation}
in all $16$ cases. It is important that the right number of $k=0$
factors appears so that there are $4$ minimal models altogether
(exceptionally, $6$ in the first two Gepner models).

In addition to the $16$ known $K3$ Gepner models we found that 
\begin{equation}
  \tK[(2,3,3,3,3,3,3)]_{U(1)^7\times \GSO}^\ast\Big(SU(2)^7;\C\Big) 
  = 
  K^\ast\big(K3;\C\big)
  \,,  
\end{equation}
as well. Although it has not the conventional number of factors, this
$\overline{k}=(0,1,1,1,1,1,1)$ Gepner model seems to yield yet another
$K3$ CFT.

There is yet another combination of levels such that the total central
charge $c=6$, which is $\overline{k}=(0,1,1,1,2,2)$. One can easily
compute that 
\begin{equation}
  \tK[(2,3,3,3,4,4)]_{U(1)^6\times \GSO}^\ast\Big(SU(2)^6;\C\Big) = 
  \left\{
    \begin{array}{cc}
      \C^{8}
      \,,& 
      \ast=0
      \\
      \C^{8}
      \,,& 
      \ast=1
    \end{array}
  \right\}
  = 
  K^\ast\big(T^4;\C\big)
  \,.  
\end{equation}
Clearly, this Gepner model describes a $T^4$ compactification with
(accidentally) enhanced $\mathcal{N}=8$ space-time supersymmetry.

\section{Calabi-Yau Threefold Gepner Models}
\label{sec:tableCY3}

First, note that a proper Calabi-Yau threefold $X$, that is a compact
K\"ahler manifold of holonomy $SU(3)$ satisfies
\begin{equation}
  \label{eq:Khodge}
  \rank K^0(X) = 2 h^{11}(X) + 2
  \,, \qquad
  \rank K^1(X) = 2 h^{21}(X) + 2
  \,.
\end{equation}
We can check this formula against the known list~\cite{Lynker:1990gh,
  Fuchs:1989yv} of $168$ Gepner models with central charge $c=9$,
which are associated to Calabi-Yau threefolds. The list of all Gepner
models is reproduced in Table~\ref{tab:CY}. If one uses these
$N=(2,2)$ SCFTs as the compactification of the $E_8\times E_8$
heterotic string, then their low-energy spectrum consists of a number
$n_{\overline{\mathbf{27}}}=h^{11}(X)$ of matter fields transforming
in the $\overline{\mathbf{27}}$ and $n_{\mathbf{27}}=h^{21}(X)$ of
field in the $\mathbf{27}$ representation of $E_6$.
\begin{table}[p]
  \centering
  {
  \tiny
  \renewcommand{\arraystretch}{0.9}
 \fontfamily{ptm}\fontseries{b}\fontshape{it}\fontsize{6pt}{8pt}
  \begin{tabular}{c|cccc}
    $\overline{k}=(k_1,k_2,\dots)$ &
    $n_{\overline{\mathbf{27}}}$ &
    $n_{\mathbf{27}}$ &
    $\rk K^1$ &
    $\rk K^0$
    \\ \hline
    $(1, 1, 1, 1, 1, 1, 1, 1, 1) $ & $ 0$ & $ 84$ & $ 2$ & $ 170
           $\\ $(1, 1, 1, 1, 1, 1, 1, 4, 0)$ & $ 0$ & $ 84$ & $ 2$ & $ 170
          $\\ $(1, 1, 1, 1, 1, 1, 2, 2, 0)$ & $ 21$ & $ 21$ & $ 48$ & $ 48
             $\\ $(1, 1, 1, 1, 1, 2, 10)$,& $2$ & $ 62$ & $ 6$ & $ 126
              $\\ $(1, 1, 1, 1, 1, 4, 4)$ & $ 1$ & $ 73$ & $ 4$ & $ 148
             $\\ $(1, 1, 1, 1, 2, 2, 4)$ & $ 11$ & $ 35$ & $ 24$ & $ 72
             $\\ $(1, 1, 1, 2, 2, 2, 2)$ & $ 21$ & $ 21$ & $ 48$ & $ 48
             $\\ $(1, 1, 1, 1, 5, 40, 0)$ & $ 23$ & $ 47$ & $ 48$ & $ 96
            $\\ $(1, 1, 1, 1, 6, 22, 0)$ & $ 16$ & $ 52$ & $ 34$ & $ 106
             $\\ $(1, 1, 1, 1, 7, 16, 0)$ & $ 8$ & $ 68$ & $ 18$ & $ 138
             $\\ $(1, 1, 1, 1, 8, 13, 0)$ & $ 17$ & $ 41$ & $ 36$ & $ 84
            $\\ $(1, 1, 1, 1, 10, 10, 0)$ & $ 7$ & $ 79$ & $ 16$ & $ 160
             $\\ $(1, 1, 1, 2, 3, 18, 0)$ & $ 21$ & $ 21$ & $ 48$ & $ 48
             $\\ $(1, 1, 1, 2, 4, 10, 0)$ & $ 2$ & $ 62$ & $ 6$ & $ 126
             $\\ $(1, 1, 1, 2, 6, 6, 0)$ & $ 21$ & $ 21$ & $ 48$ & $ 48
             $\\ $(1, 1, 1, 3, 3, 8, 0)$ & $ 21$ & $ 21$ & $ 48$ & $ 48
              $\\ $(1, 1, 1, 4, 4, 4, 0)$ & $ 0$ & $ 84$ & $ 2$ & $ 170
             $\\ $(1, 1, 2, 2, 2, 10, 0)$ & $ 10$ & $ 46$ & $ 22$ & $ 94
              $\\ $(1, 1, 2, 2, 4, 4, 0)$ & $ 3$ & $ 51$ & $ 8$ & $ 104
              $\\ $(1, 2, 2, 2, 2, 4, 0)$ & $ 1$ & $ 61$ & $ 4$ & $ 124
              $\\ $(2, 2, 2, 2, 2, 2, 0)$ & $ 0$ & $ 90$ & $ 2$ & $ 182
              $\\ $(1, 1, 2, 11, 154)$ & $71$ & $ 71$ & $ 144$ & $ 144
               $\\ $(1, 1, 2, 12, 82)$ & $ 40$ & $ 76$ & $ 82$ & $ 154
               $\\ $(1, 1, 2, 13, 58)$ & $ 26$ & $ 86$ & $ 54$ & $ 174
               $\\ $(1, 1, 2, 14, 46)$ & $ 26$ & $ 86$ & $ 54$ & $ 174
              $\\ $(1, 1, 2, 16, 34)$ & $ 16$ & $ 100$ & $ 34$ & $ 202
               $\\ $(1, 1, 2, 18, 28)$ & $ 31$ & $ 55$ & $ 64$ & $ 112
               $\\ $(1, 1, 2, 19, 26)$ & $ 41$ & $ 41$ & $ 84$ & $ 84
              $\\ $(1, 1, 2, 22, 22)$ & $ 11$ & $ 131$ & $ 24$ & $ 264
              $\\ $(1, 1, 3, 6, 118)$ & $ 55$ & $ 55$ & $ 112$ & $ 112
               $\\ $(1, 1, 3, 7, 43)$ & $ 19$ & $ 67$ & $ 40$ & $ 136
               $\\ $(1, 1, 3, 8, 28)$ & $ 19$ & $ 69$ & $ 40$ & $ 140
               $\\ $(1, 1, 3, 10, 18)$ & $ 31$ & $ 31$ & $ 64$ & $ 64
               $\\ $(1, 1, 3, 13, 13)$ & $ 7$ & $ 103$ & $ 16$ & $ 208
               $\\ $(1, 1, 4, 5, 40)$ & $ 17$ & $ 65$ & $ 36$ & $ 132
               $\\ $(1, 1, 4, 6, 22)$ & $ 10$ & $ 70$ & $ 22$ & $ 142
                $\\ $(1, 1, 4, 7, 16)$ & $ 7$ & $ 79$ & $ 16$ & $ 160
                $\\ $(1, 1, 4, 8, 13)$ & $ 12$ & $ 48$ & $ 26$ & $ 98
               $\\ $(1, 1, 4, 10, 10)$ & $ 5$ & $ 101$ & $ 12$ & $ 204
               $\\ $(1, 1, 5, 5, 19)$ & $ 17$ & $ 65$ & $ 36$ & $ 132
                $\\ $(1, 1, 6, 6, 10)$ & $ 19$ & $ 43$ & $ 40$ & $ 88
                $\\ $(1, 1, 7, 7, 7)$ & $ 4$ & $ 112$ & $ 10$ & $ 226
                $\\ $(1, 2, 2, 5, 40)$ & $ 35$ & $ 35$ & $ 72$ & $ 72
                $\\ $(1, 2, 2, 6, 22)$ & $ 8$ & $ 68$ & $ 18$ & $ 138
                $\\ $(1, 2, 2, 7, 16)$ & $ 19$ & $ 43$ & $ 40$ & $ 88
                $\\ $(1, 2, 2, 8, 13)$ & $ 27$ & $ 27$ & $ 56$ & $ 56
               $\\ $(1, 2, 2, 10, 10)$ & $ 5$ & $ 89$ & $ 12$ & $ 180
                $\\ $(1, 2, 3, 3, 58)$ & $ 23$ & $ 47$ & $ 48$ & $ 96
                $\\ $(1, 2, 3, 4, 18)$ & $ 15$ & $ 39$ & $ 32$ & $ 80
                $\\ $(1, 2, 4, 4, 10)$ & $ 2$ & $ 74$ & $ 6$ & $ 150
                $\\ $(1, 2, 4, 6, 6)$ & $ 7$ & $ 55$ & $ 16$ & $ 112
                $\\ $(1, 3, 3, 3, 13)$ & $ 3$ & $ 75$ & $ 8$ & $ 152
                $\\ $(1, 3, 3, 4, 8)$ & $ 15$ & $ 39$ & $ 32$ & $ 80
                $\\ $(1, 4, 4, 4, 4)$ & $ 1$ & $ 103$ & $ 4$ & $ 208
                $\\ $(2, 2, 2, 3, 18)$ & $ 5$ & $ 65$ & $ 12$ & $ 132
                $\\ $(2, 2, 2, 4, 10)$ & $ 3$ & $ 69$ & $ 8$ & $ 140
                 $\\ $(2, 2, 2, 6, 6)$ & $ 2$ & $ 86$ & $ 6$ & $ 174
                $\\ $(2, 2, 3, 3, 8)$ & $ 15$ & $ 39$ & $ 32$ & $ 80
                $\\ $(2, 2, 4, 4, 4)$ & $ 6$ & $ 60$ & $ 14$ & $ 122
                $\\ $(3, 3, 3, 3, 3)$ & $ 1$ & $ 101$ & $ 4$ & $ 204
            $\\ $(0, 1, 5, 41, 1804)$ & $ 251$ & $ 251$ & $ 504$ & $ 504
             $\\ $(0, 1, 5, 42, 922)$ & $ 137$ & $ 257$ & $ 276$ & $ 516
             $\\ $(0, 1, 5, 43, 628)$ & $ 95$ & $ 263$ & $ 192$ & $ 528
             $\\ $(0, 1, 5, 44, 481)$ & $ 143$ & $ 143$ & $ 288$ & $ 288
              $\\ $(0, 1, 5, 46, 334)$ & $ 47$ & $ 287$ & $ 96$ & $ 576
              $\\ $(0, 1, 5, 47, 292)$ & $ 47$ & $ 287$ & $ 96$ & $ 576
             $\\ $(0, 1, 5, 49, 236)$ & $ 107$ & $ 107$ & $ 216$ & $ 216
             $\\ $(0, 1, 5, 52, 187)$ & $ 53$ & $ 173$ & $ 108$ & $ 348
              $\\ $(0, 1, 5, 54, 166)$ & $ 23$ & $ 335$ & $ 48$ & $ 672
             $\\ $(0, 1, 5, 58, 138)$ & $ 59$ & $ 131$ & $ 120$ & $ 264
              $\\ $(0, 1, 5, 61, 124)$ & $ 17$ & $ 377$ & $ 36$ & $ 756
              $\\ $(0, 1, 5, 68, 103)$ & $ 29$ & $ 221$ & $ 60$ & $ 444
              $\\ $(0, 1, 5, 76, 89)$ & $ 83$ & $ 83$ & $ 168$ & $ 168
              $\\ $(0, 1, 5, 82, 82)$ & $ 11$ & $ 491$ & $ 24$ & $ 984
             $\\ $(0, 1, 6, 23, 598)$ & $ 119$ & $ 167$ & $ 240$ & $ 336
             $\\ $(0, 1, 6, 24, 310)$ & $ 66$ & $ 174$ & $ 134$ & $ 350
              $\\ $(0, 1, 6, 25, 214)$ & $ 48$ & $ 180$ & $ 98$ & $ 362
              $\\ $(0, 1, 6, 26, 166)$ & $ 34$ & $ 190$ & $ 70$ & $ 382
              $\\ $(0, 1, 6, 28, 118)$ & $ 24$ & $ 204$ & $ 50$ & $ 410
              $\\ $(0, 1, 6, 30, 94)$ & $ 18$ & $ 222$ & $ 38$ & $ 446
              $\\ $(0, 1, 6, 31, 86)$ & $ 57$ & $ 81$ & $ 116$ & $ 164
              $\\ $(0, 1, 6, 34, 70)$ & $ 14$ & $ 242$ & $ 30$ & $ 486
              $\\ $(0, 1, 6, 38, 58)$ & $ 23$ & $ 143$ & $ 48$ & $ 288
              $\\ $(0, 1, 6, 40, 54)$ & $ 33$ & $ 105$ & $ 68$ & $ 212
              $
 \end{tabular}
 \hfill
  \begin{tabular}{c|cccc}
    $\overline{k}=(k_1,k_2,\dots)$ &
    $n_{\overline{\mathbf{27}}}$ &
    $n_{\mathbf{27}}$ &
    $\rk K^1$ &
    $\rk K^0$
    \\ \hline
                 $(0, 1, 6, 46, 46)$ & $ 9$ & $ 321$ & $ 20$ & $ 644
             $\\ $(0, 1, 7, 17, 340)$ & $ 71$ & $ 143$ & $ 144$ & $ 288
              $\\ $(0, 1, 7, 18, 178)$ & $ 42$ & $ 150$ & $ 86$ & $ 302
              $\\ $(0, 1, 7, 19, 124)$ & $ 28$ & $ 160$ & $ 58$ & $ 322
               $\\ $(0, 1, 7, 20, 97)$ & $ 45$ & $ 93$ & $ 92$ & $ 188
              $\\ $(0, 1, 7, 22, 70)$ & $ 15$ & $ 183$ & $ 32$ & $ 368
              $\\ $(0, 1, 7, 25, 52)$ & $ 10$ & $ 214$ & $ 22$ & $ 430
              $\\ $(0, 1, 7, 28, 43)$ & $ 18$ & $ 126$ & $ 38$ & $ 254
               $\\ $(0, 1, 7, 34, 34)$ & $ 7$ & $ 271$ & $ 16$ & $ 544
             $\\ $(0, 1, 8, 14, 238)$ & $ 50$ & $ 134$ & $ 102$ & $ 270
              $\\ $(0, 1, 8, 16, 88)$ & $ 17$ & $ 155$ & $ 36$ & $ 312
              $\\ $(0, 1, 8, 18, 58)$ & $ 10$ & $ 178$ & $ 22$ & $ 358
               $\\ $(0, 1, 8, 22, 38)$ & $ 22$ & $ 82$ & $ 46$ & $ 166
               $\\ $(0, 1, 8, 28, 28)$ & $ 5$ & $ 251$ & $ 12$ & $ 504
              $\\ $(0, 1, 9, 12, 229)$ & $ 79$ & $ 79$ & $ 160$ & $ 160
              $\\ $(0, 1, 9, 13, 108)$ & $ 59$ & $ 59$ & $ 120$ & $ 120
               $\\ $(0, 1, 9, 20, 31)$ & $ 9$ & $ 129$ & $ 20$ & $ 260
             $\\ $(0, 1, 10, 11, 154)$ & $ 23$ & $ 143$ & $ 48$ & $ 288
              $\\ $(0, 1, 10, 12, 82)$ & $ 15$ & $ 147$ & $ 32$ & $ 296
              $\\ $(0, 1, 10, 13, 58)$ & $ 11$ & $ 155$ & $ 24$ & $ 312
              $\\ $(0, 1, 10, 14, 46)$ & $ 8$ & $ 164$ & $ 18$ & $ 330
              $\\ $(0, 1, 10, 16, 34)$ & $ 5$ & $ 185$ & $ 12$ & $ 372
              $\\ $(0, 1, 10, 18, 28)$ & $ 10$ & $ 106$ & $ 22$ & $ 214
              $\\ $(0, 1, 10, 19, 26)$ & $ 16$ & $ 76$ & $ 34$ & $ 154
               $\\ $(0, 1, 10, 22, 22)$ & $ 3$ & $ 243$ & $ 8$ & $ 488
              $\\ $(0, 1, 11, 11, 76)$ & $ 23$ & $ 143$ & $ 48$ & $ 288
              $\\ $(0, 1, 12, 12, 40)$ & $ 6$ & $ 180$ & $ 14$ & $ 362
               $\\ $(0, 1, 12, 13, 33)$ & $ 43$ & $ 43$ & $ 88$ & $ 88
              $\\ $(0, 1, 12, 19, 19)$ & $ 7$ & $ 151$ & $ 16$ & $ 304
              $\\ $(0, 1, 13, 13, 28)$ & $ 4$ & $ 208$ & $ 10$ & $ 418
              $\\ $(0, 1, 13, 18, 18)$ & $ 11$ & $ 107$ & $ 24$ & $ 216
              $\\ $(0, 1, 14, 14, 22)$ & $ 7$ & $ 127$ & $ 16$ & $ 256
               $\\ $(0, 1, 16, 16, 16)$ & $ 2$ & $ 272$ & $ 6$ & $ 546
             $\\ $(0, 2, 3, 19, 418)$ & $ 119$ & $ 119$ & $ 240$ & $ 240
             $\\ $(0, 2, 3, 20, 218)$ & $ 65$ & $ 125$ & $ 132$ & $ 252
              $\\ $(0, 2, 3, 22, 118)$ & $ 33$ & $ 141$ & $ 68$ & $ 284
              $\\ $(0, 2, 3, 23, 98)$ & $ 33$ & $ 141$ & $ 68$ & $ 284
               $\\ $(0, 2, 3, 26, 68)$ & $ 39$ & $ 87$ & $ 80$ & $ 176
              $\\ $(0, 2, 3, 28, 58)$ & $ 17$ & $ 173$ & $ 36$ & $ 348
              $\\ $(0, 2, 3, 34, 43)$ & $ 55$ & $ 55$ & $ 112$ & $ 112
              $\\ $(0, 2, 3, 38, 38)$ & $ 11$ & $ 227$ & $ 24$ & $ 456
              $\\ $(0, 2, 4, 11, 154)$ & $ 53$ & $ 89$ & $ 108$ & $ 180
               $\\ $(0, 2, 4, 12, 82)$ & $ 30$ & $ 96$ & $ 62$ & $ 194
              $\\ $(0, 2, 4, 13, 58)$ & $ 20$ & $ 104$ & $ 42$ & $ 210
              $\\ $(0, 2, 4, 14, 46)$ & $ 16$ & $ 112$ & $ 34$ & $ 226
              $\\ $(0, 2, 4, 16, 34)$ & $ 12$ & $ 126$ & $ 26$ & $ 254
               $\\ $(0, 2, 4, 18, 28)$ & $ 20$ & $ 74$ & $ 42$ & $ 150
               $\\ $(0, 2, 4, 19, 26)$ & $ 28$ & $ 52$ & $ 58$ & $ 106
               $\\ $(0, 2, 4, 22, 22)$ & $ 8$ & $ 164$ & $ 18$ & $ 330
               $\\ $(0, 2, 5, 8, 138)$ & $ 44$ & $ 80$ & $ 90$ & $ 162
               $\\ $(0, 2, 5, 10, 40)$ & $ 23$ & $ 59$ & $ 48$ & $ 120
               $\\ $(0, 2, 5, 12, 26)$ & $ 8$ & $ 116$ & $ 18$ & $ 234
               $\\ $(0, 2, 6, 7, 70)$ & $ 19$ & $ 91$ & $ 40$ & $ 184
               $\\ $(0, 2, 6, 8, 38)$ & $ 12$ & $ 96$ & $ 26$ & $ 194
               $\\ $(0, 2, 6, 10, 22)$ & $ 6$ & $ 114$ & $ 14$ & $ 230
               $\\ $(0, 2, 6, 14, 14)$ & $ 4$ & $ 148$ & $ 10$ & $ 298
               $\\ $(0, 2, 7, 7, 34)$ & $ 19$ & $ 91$ & $ 40$ & $ 184
               $\\ $(0, 2, 7, 10, 16)$ & $ 10$ & $ 70$ & $ 22$ & $ 142
               $\\ $(0, 2, 8, 8, 18)$ & $ 6$ & $ 120$ & $ 14$ & $ 242
               $\\ $(0, 2, 8, 10, 13)$ & $ 18$ & $ 42$ & $ 38$ & $ 86
               $\\ $(0, 2, 10, 10, 10)$ & $ 3$ & $ 165$ & $ 8$ & $ 332
               $\\ $(0, 3, 3, 9, 108)$ & $ 39$ & $ 79$ & $ 80$ & $ 160
               $\\ $(0, 3, 3, 10, 58)$ & $ 25$ & $ 85$ & $ 52$ & $ 172
               $\\ $(0, 3, 3, 12, 33)$ & $ 27$ & $ 59$ & $ 56$ & $ 120
               $\\ $(0, 3, 3, 13, 28)$ & $ 9$ & $ 117$ & $ 20$ & $ 236
               $\\ $(0, 3, 3, 18, 18)$ & $ 7$ & $ 143$ & $ 16$ & $ 288
               $\\ $(0, 3, 4, 6, 118)$ & $ 33$ & $ 69$ & $ 68$ & $ 140
               $\\ $(0, 3, 4, 7, 43)$ & $ 19$ & $ 67$ & $ 40$ & $ 136
                $\\ $(0, 3, 4, 8, 28)$ & $ 7$ & $ 91$ & $ 16$ & $ 184
               $\\ $(0, 3, 4, 10, 18)$ & $ 13$ & $ 49$ & $ 28$ & $ 100
               $\\ $(0, 3, 4, 13, 13)$ & $ 7$ & $ 103$ & $ 16$ & $ 208
               $\\ $(0, 3, 5, 5, 68)$ & $ 23$ & $ 71$ & $ 48$ & $ 144
                $\\ $(0, 3, 6, 6, 18)$ & $ 7$ & $ 63$ & $ 16$ & $ 128
                $\\ $(0, 3, 8, 8, 8)$ & $ 1$ & $ 145$ & $ 4$ & $ 292
                $\\ $(0, 4, 4, 5, 40)$ & $ 8$ & $ 86$ & $ 18$ & $ 174
                $\\ $(0, 4, 4, 6, 22)$ & $ 6$ & $ 90$ & $ 14$ & $ 182
                $\\ $(0, 4, 4, 7, 16)$ & $ 3$ & $ 99$ & $ 8$ & $ 200
                $\\ $(0, 4, 4, 8, 13)$ & $ 7$ & $ 61$ & $ 16$ & $ 124
               $\\ $(0, 4, 4, 10, 10)$ & $ 2$ & $ 128$ & $ 6$ & $ 258
               $\\ $(0, 4, 5, 5, 19)$ & $ 17$ & $ 65$ & $ 36$ & $ 132
                $\\ $(0, 4, 6, 6, 10)$ & $ 6$ & $ 66$ & $ 14$ & $ 134
                $\\ $(0, 4, 7, 7, 7)$ & $ 4$ & $ 112$ & $ 10$ & $ 226
                $\\ $(0, 5, 5, 5, 12)$ & $ 2$ & $ 122$ & $ 6$ & $ 246
                $\\ $(0, 6, 6, 6, 6)$ & $ 1$ & $ 149$ & $ 4$ & $ 300$
 \end{tabular}
 }
  \caption{$c=9$ Gepner models.}
  \label{tab:CY}
\end{table}

One can check that the formula eq.~\eqref{eq:Khodge} is obeyed for
each Gepner model except for the $7$ cases with
$n_{\overline{\mathbf{27}}}=n_{\mathbf{27}}=21$. The obvious
explanation is that this misfit is associated $K3 \times T^2$, which
has Hodge numbers
\begin{equation}
  h^{pq}(K3 \times T^2)=~
  \vcenter{\xymatrix@!0@=7mm@ur{
    1 &  1 &  1 & 1 \\
    1 & 21 & 21 & 1 \\
    1 & 21 & 21 & 1 \\
    1 &  1 &  1 & 1 
  }}
  \,.
\end{equation}
Since $K3 \times T^2$ has only $SU(2)$ holonomy, that is, it is not a
proper Calabi-Yau manifold, it does not have to obey
eq.~\eqref{eq:Khodge}. Adding up the even and odd cohomology groups,
we find that 
\begin{equation}
  K^0\big(K3\times T^2\big) =
  \Z^{48}
  \,, \qquad
  K^1\big(K3\times T^2\big) =
  \Z^{48}
  \,.
\end{equation}
These topological K-groups are in precise agreement with what we
computed using the coset eq.~\eqref{eq:Kgepner}.

\section*{Acknowledgments}

This research was supported in part by the Department of Physics and
the Math/Physics Research Group at the University of Pennsylvania
under cooperative research agreement DE-FG02-95ER40893 with the
U.~S.~Department of Energy and an NSF Focused Research Grant
DMS0139799 for ``The Geometry of Superstrings.''. This work was
partially supported by the DFG, DAAD, and European RTN Program
MRTN-CT-2004-503369.

We would like to acknowledge useful discussions with Volker Schomerus
and Katrin Wendland.

\bibliographystyle{JHEP} \renewcommand{\refname}{Bibliography}
\addcontentsline{toc}{section}{Bibliography} \bibliography{main}

\end{document}

%% file: S1A.pstex_t
\begin{picture}(0,0)%
\includegraphics{S1A.pstex}%
\end{picture}%
\setlength{\unitlength}{4144sp}%
\begingroup\makeatletter\ifx\SetFigFont\undefined%
\gdef\SetFigFont#1#2#3#4#5{%
  \reset@font\fontsize{#1}{#2pt}%
  \fontfamily{#3}\fontseries{#4}\fontshape{#5}%
  \selectfont}%
\fi\endgroup%
\begin{picture}(5007,2463)(-252,-8173)
\put(901,-6271){\makebox(0,0)[b]{\smash{{\SetFigFont{12}{14.4}{\rmdefault}{\mddefault}{\updefault}{\color[rgb]{0,0,0}$U(1)$ orbits}%
}}}}
\put(2836,-5866){\makebox(0,0)[b]{\smash{{\SetFigFont{12}{14.4}{\rmdefault}{\mddefault}{\updefault}{\color[rgb]{0,0,0}$SU(2)\simeq S^3$}%
}}}}
\put(1576,-7306){\makebox(0,0)[lb]{\smash{{\SetFigFont{12}{14.4}{\rmdefault}{\mddefault}{\updefault}{\color[rgb]{1,0,0}$D^2=SU(2)/U(1)$}%
}}}}
\end{picture}%

%% file: S1B.pstex_t
\begin{picture}(0,0)%
\includegraphics{S1B.pstex}%
\end{picture}%
\setlength{\unitlength}{4144sp}%
\begingroup\makeatletter\ifx\SetFigFont\undefined%
\gdef\SetFigFont#1#2#3#4#5{%
  \reset@font\fontsize{#1}{#2pt}%
  \fontfamily{#3}\fontseries{#4}\fontshape{#5}%
  \selectfont}%
\fi\endgroup%
\begin{picture}(3162,3119)(901,-7081)
\put(3591,-4554){\makebox(0,0)[lb]{\smash{{\SetFigFont{12}{14.4}{\rmdefault}{\mddefault}{\updefault}{\color[rgb]{0,0,0}$\Z_5$ action}%
}}}}
\put(1416,-6438){\makebox(0,0)[lb]{\smash{{\SetFigFont{12}{14.4}{\rmdefault}{\mddefault}{\updefault}{\color[rgb]{1,0,0}$D^2=SU(2)/U(1)$}%
}}}}
\put(1857,-5934){\makebox(0,0)[b]{\smash{{\SetFigFont{12}{14.4}{\rmdefault}{\mddefault}{\updefault}{\color[rgb]{0,0,0}$U(1)$ orbits}%
}}}}
\put(1857,-6114){\makebox(0,0)[b]{\smash{{\SetFigFont{12}{14.4}{\rmdefault}{\mddefault}{\updefault}{\color[rgb]{0,0,0}vertical}%
}}}}
\end{picture}%